\documentstyle[proceedings,numreferences,graphicx,amssymb]{crckapb} 

\def\be{\begin{equation}}
\def\ee{\end{equation}}
\def\bea{\begin{eqnarray}}
\def\eea{\end{eqnarray}}

\begin{opening}
\title{HADRONIC TRANSPORT MODEL\protect\\ WITH A PHASE
TRANSITION$^*$}
\author{P.\ Danielewicz$\,\,$}
\author{$\!\!\!\!\!\!^1\,\,$ P.-B.\ Gossiaux$^{1,2}$}
\author{R.\ A.\ Lacey$^3$}
\institute{ \\[-1.5ex]
$^1$National Superconducting Cyclotron Laboratory\\ and
the Department of Physics and Astronomy, Michigan State University,
East Lansing, MI 48824-1321, USA \\[1.5ex]
\hspace*{0em}$^2$SUBATECH, Ecole des Mines, F-44070 Nantes, France \\[1.5ex]
\hspace*{0em}$^3$Departments of Chemistry and Physics, State University of
New York at Stony Brook,
Stony Brook, NY 11794-3400, USA }
\end{opening}
\runningtitle{TRANSPORT MODEL WITH PHASE TRANSITION}

\begin{document}

\begin{abstract}
{
We specify
a~tractable transport model with
thermodynamic properties close to those expected for the
strongly
interacting matter.  In particular, at high temperatures,
the~matter undergoes a~phase transition, such as to the
quark-gluon plasma, with a~drop in masses of elementary
excitations and a~rapid increase in the number of degrees of
freedom.
We show that a~softening of the equation of
state such
as associated with the transition to quark-gluon plasma should
be observable in the elliptic-flow excitation
function from heavy-ion reactions.
}
\end{abstract}

\section{Introduction}
One of the important goals of the heavy-ion reaction
studies is the detection of quark-gluon (QG)
plasma.\footnote{$^*$Talk given at the
Workshop on Nuclear Matter
in Different Phases and Transitions, Les Houches, March~31 --
April~10, 1998.}
Reaching the transition to QG plasma requires
a~significant increase in hadron density in a~reaction,
through an~increase in baryon density, or an~increase in
temperature, or both.
General expectations concerning the approach to the transition
are as follows.  When hadrons increase in density in
some spatial
region, they push out the standard nonperturbative vacuum.
The~fraction of the volume taken by the nonperturbative
vacuum decreases and this, on the average, reduces hadron
masses associated with the condensate in that
vacuum.  At~the transition, the~masses of elementary
excitations, now quarks and gluons, drop to values close to
zero.  The~number of the degrees of freedom at the phase
transition dramatically increases.

Specific quantitative information on the transition to QG
plasma comes
from numerical quantum-chromodynamic lattice calculations.
Since these results pertain to a~baryonless system at
equilibrium, though, they are insufficient for the {\em
reaction description}.
The~relatively well understood domain is that of the strongly
interacting matter
at low energy densities.  That matter is describable in terms
of individual
hadrons scattering on each other, with cross sections close to
those in free space. The~hadrons further feel the overall mean
field produced
by remaining hadrons in the vicinity.  Transport theory based
on such concepts had much success with low and intermediate
energy reactions.  Ground-state properties of nuclear matter
are rather well known.  The~areas of most uncertainty regarding
the strongly-interacting matter include the QG plasma out of
equilibrium.  The~conversion of the plasma into hadrons is
not comprehended.  One can suspect that somehow the
characteristic hadronic distances and time-scales are involved.

The need to test for the presence of the phase transition in
reactions and the difficult theoretic situation described above
lead us
to consider a~dynamic hadronic model for reactions, consistent
with all
known limits such as low-density hadron matter and
the thermal
equilibrium at baryon chemical potential $\mu = 0$.  The~model
could be, otherwise, applied in general nonequilibrium
situations.
In the simplest possible model, the~particle masses would be
reduced by one common factor in
connection with the phase transition:
\be
m_0 \hspace{1em} \rightarrow \hspace{1em} m = m_0 \, S \, .
\ee
The mass reduction factor $S$ should tend to zero as particle
density increases.  Physically, as the particle density
increases, the~particles
present in a~certain region may start to overlap with each
other.  Overcounting of the degrees of freedom could be treated
in terms of excluded volume nonrelativistically, but no similar
simple and consistent procedure exists relativistically.
Relying on the fact that the number of effective degrees of
freedom should not exceed the number of fundamental degrees of
freedom, we decided to adopt a~cutoff in the mass spectrum of
the included hadrons to roughly match the number of quarks and
gluons, and disregard the excluded volume.  Thus, the number of
quarks and gluons is $24 +
16 =40$.  We include nucleons, deltas, and their antiparticles,
pions and $\rho$ mesons.  When these particles become massless,
we roughly match the number of massless degrees of freedom in
the plasma, with $8 + 32 + 3 + 9 = 52$, expecting a~sensible
increase in the entropy and in other thermodynamic quantities
across the transition.

For the selected degrees of freedom, one needs to specify
a~dynamics.  At low energies, the~combination of collisions and
mean field was successful; the~mean field could be used to
lower masses.  The~common approach is to start from
a~Lagrangian and adopt a~mean-field approximation to resulting
equations of motion.  The~mean field often then has undesirable
properties and, to repair these, more and more nonlinear terms
are added the Lagrangian making it cumbersome.  We decided to
cirmcumvent these steps by formulating our approach within the
relativistic Landau theory.

\section{Relativistic Landau Theory}

Within the Landau theory, the~interactions are specified by
giving the energy density as a~functional of
particle phase-space distributions~$f$~\cite{bay76}
\be
T^{00} = e \equiv e \lbrace f \rbrace \, .
\label{T00}
\ee
The single-particle energies represent functional derivatives
of the energy
\be
\epsilon_{\bf p}^i = {\delta e \over \delta f^i ({\bf p},
{\bf r}, t)}  \, ,
\ee
where $i$ is the particle index.  The~single-particle energy
and momentum, $({\bf p}, \epsilon_{\bf p})$, transform,
generally, as a~four-vector.

In the simplest parametrization of the energy, ensuring
covariance, the~net energy consists of the sum of kinetic
energies and corrections for interactions dependent on
scalar and vector densities:
\be
e = \sum_i \int d{\bf p} \, \epsilon_{\bf p}^i \,
 f^i ({\bf p})
+ e_s(\rho_s) + e_v(\rho_v) \, ,
\ee
in a~local frame, with
\be
\rho_s = \sum_i \int d {\bf p} \, {m^i \, m_0^i
\over \sqrt{m^{i2} + p^2}} \, f^i({\bf p})
\ee
and
\be
\rho_v = \sum_i B^i \int d {\bf p} \, f^i({\bf p}) \, .
\ee
Dependence on two densities, $\rho_s$ and $\rho_v$, is needed
to parametrize, independently, the thermodynamic properties
along the $\mu = 0$ and $T = 0$ axes.

Contributions of different hadrons to the scalar density are
weighted with the hadron mass, to ensure that the masses change
by the common factor~$S$,
\be
m^i = m_0^i \, S(\rho_s) \, ,
\hspace{1em} \mbox{where} \hspace{1em}
S = \int {d \rho_s \over \rho_s} \, {d e_s \over
d \rho_s} \, .
\ee
The single-particle energy in a~local rest frame is then
\be
\epsilon_{\bf p}^i = \sqrt{ m^{i2} + p^2} +
B^i \, V(\rho_v) \, , \hspace{.5em}
\mbox{with} \hspace{.5em}
V = \int {d \rho_v \over \rho_v} \, {d e_v \over
d \rho_v} \, .
\ee
In any frame, the baryon four-current is
\be
\rho_v^\mu = \left( \sum_i \int d{\bf p} \, f^i,
\sum_i \int d{\bf p} \, {\partial \epsilon^i \over \partial
{\bf p}} \, f^i \right) \, ,
\ee
and $\rho_v^\mu \, \rho_{v \mu} = \rho_v^2$.
The~canonical four-momentum $p^\mu$ may be expressed, similarly
to electrodynamics, in terms
of the kinematic four-momentum $p^{* \mu}$ and the four-vector
potential in the direction of the
four-current,
\be
p^{i\mu} = p^{i * \mu} + B^i \, V \, \rho_v^\mu/ \rho_v \, ,
\ee
with $p^{i*2} = m^{i2}$.
Locally, the canonical and kinematic three-momenta are
identical, ${\bf p}^{i*} = {\bf p}^i$.

Now we move on to the thermodynamic properties of matter.
Our results for equilibrium generalize those
found within the Walecka model.

\section{Thermodynamic Properties}

The transition to QG plasma is characterized by an increase in
the number of the degrees of freedom and by a~decrease in the
masses of elementary quanta.  In the discussed model,
the~transition may be produced by requiring the drop of masses
with
an~increase in density.  The~decrease in masses
should lead to
an~additional increase in the number of particles present at
a~given~$T$
and, in turn, to an~additional decrease in the masses.
Eventually, as~$T$
grows the system may become unstable and a~phase transition
can take place.

\begin{figure}
\centerline{\includegraphics[angle=0,
width=.63\linewidth]{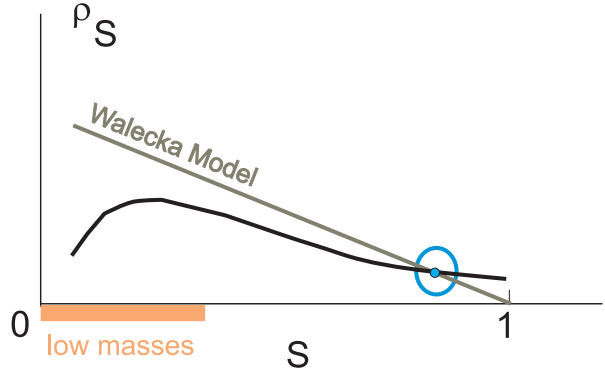}}

\caption{The values of $\rho_s$ and $S$ at a~given $T$ are
found in the $\rho_s-S$ plot from crossing of the lines
given by the $S(\rho_s)$ dependence (straight
line in the Walecka model) and by the consistency relation
(line with the hump).}
\label{rhoS}
\end{figure}

\begin{figure}
\centerline{\includegraphics[angle=0,
width=.75\linewidth]{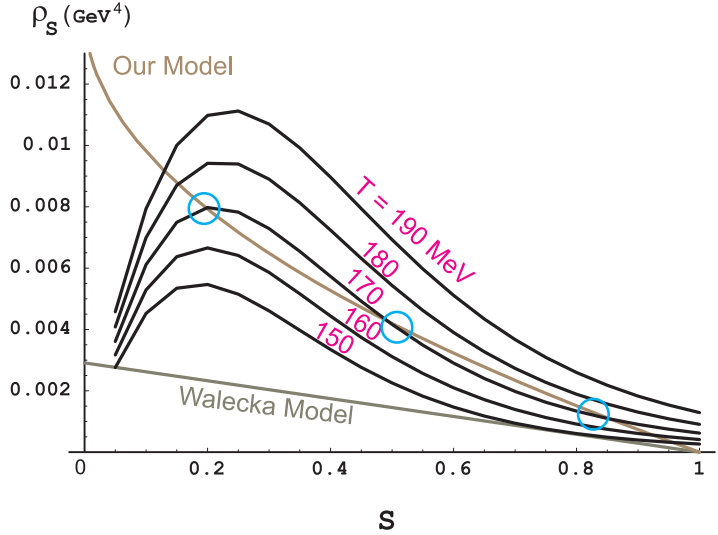}}

\caption{Three crossings of the $S(\rho_s)$ line with
the $\rho_s (S,T)$ line indicate the presence of a~phase
transition.}

\label{rhoST}
\end{figure}

\begin{figure}
\centerline{\includegraphics[angle=0,
width=.82\linewidth]{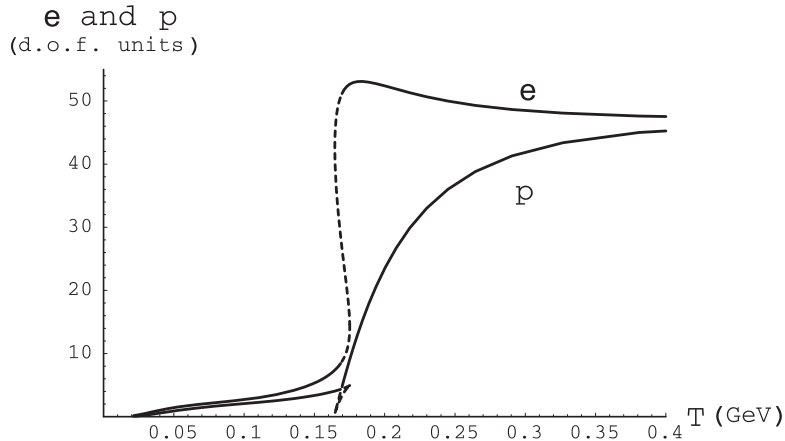}}

\caption{Energy density $e$ and pressure $p$, scaled by the
factors $\pi^2 \, T^4/30$ and $\pi^2 \, T^4/90$, respectively,
as a~function of~$T$.}

\label{ep}
\end{figure}

In asessing whether or not the phase transition takes place, it
is first necessary to determine what mass reduction is reached
at any~$T$.  In the model, the dependence $S(\rho_s)$ is
prescribed; at low~$\rho_s$, $S \simeq 1 - a \, \rho_s$,
with $a>0$, given the considerations before.  Besides,
the~consistency condition, from the definition of the density,
must be met at a~given~$T$:
\bea
\nonumber
\rho_s & \equiv & \rho_s(S,T) \\ & = & \sum_i \int d{\bf p} \,
{m_0^{i2} \,  S \over \sqrt{m_0^{i2} \, S^2 + p^2} }  \,  {1
\over \exp{\left( \sqrt{m_0^{i2} \, S^2 + p^2 }/T\right)} \pm 1}
\, ,
\label{cons}
\eea
where the equilibrium form of $f$ was inserted.
The~two equations give $\rho_s$ and $S$ for a~given~$T$, as
schematically illustrated for the Walecka model in
Fig.~\ref{rhoS} displaying
a~$\rho_s - S$ plot.  The~values are found from the crossing of
the lines given by~(\ref{cons}) and by~$S(\rho_s)$.
The~second of the dependencies is linear
in~the Walecka model
at all~$\rho_s$.
At sufficiently low $T$, only one crossing is found but, as~$T$
grows, the~hump in the curve from the consistency relation
grows.  Eventually, three crossings may be found, as shown in
Fig.~\ref{rhoST}, indicating the presence of a~phase
transition.
In the Walecka model the interactions are very strong:
\be
S = 1 - 2.6 \, (\mbox{fm}^3/\mbox{GeV}) \, \rho_s \, ,
\ee
and the phase transition takes place at low temperatures $T <
100$~MeV.  We use a~weaker dependence of $S$ on~$\rho_s$,
\be
S = \left(1 - 0.54 \, (\mbox{fm}^3/\mbox{GeV}) \, \rho_s
\right)^2 \, ,
\ee
getting the phase transition at $T \approx 170$~MeV as found
in lattice calculations~\cite{kar98}.   Figure~\ref{ep}
displays the~energy density $e$ and
pressure~$p$ as a~function of $T$ at $\mu=0$ in our model,
divided by the customary factors of $\pi^2 \, T^4
/30$ and $\pi^2 \, T^4 /90$, respectively, to show the
effective number
of degrees of freedom.  A~characteristic knot is seen in $p(T)$
indicating the transition.  Qualitative features found in
the lattice
calculations~\cite{kar98} are reproduced naturally within the
model, such as the rapid rise
of~$e$ across the transition region and a~slow rise of~$p$.
Having set the thermodynamic properties at $\mu = 0$, we
turn to the properties at~$T=0$.

\begin{figure}
\centerline{\includegraphics[angle=0,
width=.78\linewidth]{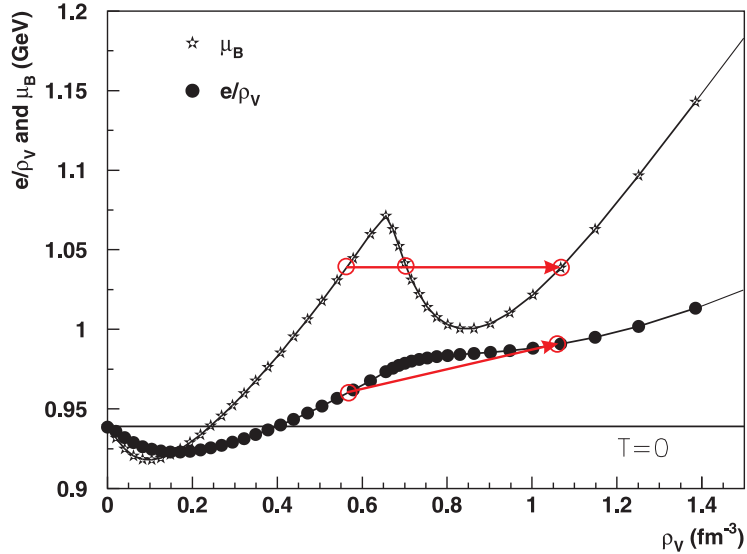}}

\caption{Chemical potential $\mu$ and energy per baryon
$e/\rho_v$ at $T=0$ as a~function of~$\rho_v$.
The~high-density phase transition is indicated with arrows.}

\label{T0}
\end{figure}

At $T=0$ the~energy per baryon $e/\rho_v$ should have a~minimum
of $939-16$~MeV, at $\rho_{v} = \rho_{0} =
0.16$~fm$^{-3}$, with the curvature characterized by the
incompressibility $K \simeq 210$~MeV.  For some
prescribed dependence
$V(\rho_v)$, the~density $\rho_v$ and the potential~$V$, at
a~given~$\mu$, may be found with the help of the
consistency relation:
\be
\rho_v \equiv \rho_v(V) = \sum_i B^i \int d{\bf p} \, \theta
\left(\mu - \sqrt{ m^{i2} + p^2} - B^i \, V \right) \, .
\label{rhoV}
\ee
The values can be found from a~crossing of the lines in
a~$\rho_v-V$ plane.  Note that the already set $S(\rho_s)$
enters into~(\ref{rhoV}).  Three rather than one crossing in
the~$\rho_v-V$ plane indicate a~phase transition.  We take
$V$ out of a~combination of powers, exclusively repulsive,
\be
V = {a \, (\rho_v/\rho_{v0})^2 \over 1 + b \, {\rho_v/\rho_{v0}
} + c \, (\rho_v/\rho_{v0})^{5/3}} \, ,
\ee
and adjust the parameters to reproduce the ground-state
nuclear-matter properties.  We find $a = 146.32$~MeV, and
$b=0.4733$.  The~value of $c=51.48$ and the power in the term
that $c$ multiplies are fixed by the requirement that
 the equation of state of a~free quark gas is
reproduced~\cite{gos98}
 at
high~$\rho_v$.  With these parameters we find a~first-order
phase transition at $T=0$, see Fig.~\ref{T0}, taking the system
from $\rho_v \sim
3.5 \, \rho_{0}$ to $\rho_v \sim 7 \, \rho_{0}$.
The~phase
transition is fragile, i.e.\ moderate changes in the
parameters replace the phase transition by a~transitional
behavior.  In~any case, though, the~matter exhibits a~rapid
change of properties along the $T=0$ axis above $3 \,
\rho_{0}$.  The~masses drop rapidly as the scalar
density increases.  This behavior reflects that along the
$\mu=0$ axis.

\begin{figure}
\centerline{\includegraphics[angle=0,
width=.79\linewidth]{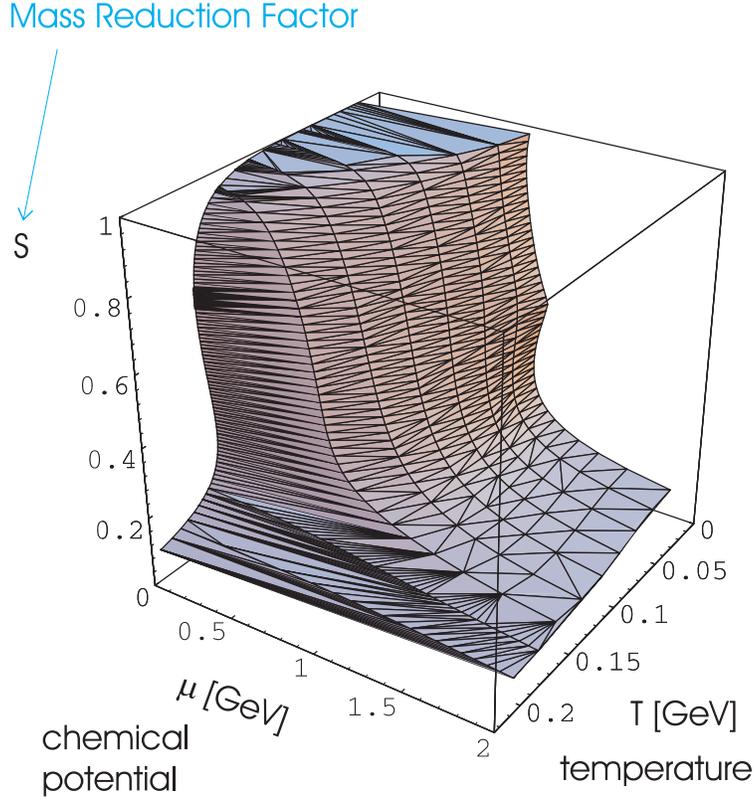}}

\caption{Mass reduction factor~$S$ as a~function of chemical
potential~$\mu$ and temperature~$T$.}

\label{SmT}
\end{figure}

When we consider all possible values of $\mu$ and~$T$, we find
the phase transitions only in the vicinity of $\mu = 0$ and
$T=0$ axes.  For moderate values of $\mu$ {\em and} $T$,
rather than going through a~phase transition,
the~matter exhibits only a~transitional behavior.
This is illustrated with
Fig.~\ref{SmT} that shows the mass reduction factor $S$ as
a~function of both $\mu$ and $T$.  At high temperatures and/or
baryon densities, the~masses fall to 20\% or less of vacuum
values.

\section{Transport Theory}

Consistently with~(\ref{T00}), the~spatial and temporal changes
in the phase-space distribution functions are described by the
Boltzmann equation
\be
{\partial f \over \partial t} + {\partial \epsilon_{\bf p} \over
\partial {\bf p} }
\, {\partial f \over \partial {\bf r}} -
{\partial \epsilon_{\bf p} \over
\partial {\bf r} } \, {\partial f \over \partial {\bf p}}
  =  I \, ,
\label{Bol}
\ee
which has the same general form relativistically as
nonrelativistically.  On the l.h.s.\ of (\ref{Bol}), $\partial
\epsilon_{\bf
p} / \partial {\bf p}$ is the velocity and $ - \partial
\epsilon_{\bf p} / \partial {\bf r}$ is force, while $I$ on the
r.h.s.\ is the collision integral.  In terms of kinematic
variables, the~Boltzmann equation acquires a~simple form:
\be
{\partial f \over \partial t} + {{\bf p}^* \over
\epsilon_{\bf p}^*  }
\, {\partial f \over \partial {\bf r}} -
{\partial \over
\partial {\bf r} }  \left( \epsilon_{\bf p}^* + V^0 \right)
\, {\partial f \over \partial {\bf p}^*}   =  I \, .
\label{Bolk}
\ee
This result generalizes the one obtained by Ko {\em et
al.}~\cite{ko87} within the Walecka model.
The~gradient in the force in (\ref{Bolk}) does not just act on
the potential, since the kinetic energy depends on position
through mass.

Collisions can, generally, change the particle number and,
thus, the integral is:
\bea
\nonumber
\gamma_1 \, I & = & \sum_{n,n' \ge 2}
\int { d{\bf p}_2 \over \gamma_2} \ldots
{ d{\bf p}_n \over \gamma_n} \,
\int { d{\bf p}_1' \over \gamma_1'} \ldots
{ d{\bf p}_{n'}' \over \gamma_{n'}'}
| {\cal M} |^2 \\
\nonumber
& & \times
\delta \left(\sum_{i'=1}^{n'} p_{i'}' - \sum_{i=1}^n p_{i} \right)
(f_1' \ldots f_{n'}' - f_1 \ldots f_n)
\\
\nonumber
 & = & \sum_{n,n' \ge 2}
\int { d{\bf p}_2^* \over \gamma_2} \ldots
{ d{\bf p}_n^* \over \gamma_n} \,
\int { d{\bf p}_1^{*'} \over \gamma_1'} \ldots
{ d{\bf p}_{n'}^{*'} \over \gamma_{n'}'}
| {\cal M} |^2 \\ & & \times
\delta \left(\sum_{i'=1}^{n'} p_{i'}^{*'} - \sum_{i=1}^n p_{i}^*
\right) (f_1' \ldots f_{n'}' - f_1 \ldots f_n) \,
\label{I=}
\eea
where $n$ and $n'$ are particle numbers in the initial and
final states.  The~rate for collisions is proportional to
the respective matrix-element squared, energy and momentum are
conserved in the collisions, and the rate of change in the
occupation results from the difference in gain and loss.
The~statistics is suppressed in~(\ref{I=}).  Since, locally,
canonical
momenta differ from kinematic momenta by a~constant shift,
the~collision integral acquires a~particular simple form in the
kinematic variables~(\ref{I=}).  Compared to vacuum, in the
medium the~mass scale just changes by the factor
of~$S$.

When aiming at a~certain equation of state~(EOS) in
a~calculation, it
is essential to obey detailed balance relations for elementary
collision processes.  That
is relatively straightforward for processes with at most two
particles in the initial and final states, but difficult for
processes with more particles.  Given this, we adopt
a~compromise in our model, treating high-
and low-energy processes differently.
The~elementary low-energy processes, that establish
thermodynamic equilibrium, with at~most two particles in any
state,
have a~strictly enforced detailed balance.  This is in contrast
to the high-energy
processes for which the inverse processes are less likely.
The~high-energy
production processes are parametrized using
experimental data
on net cross sections, pion multiplicities, hadron rapidity,
and transverse-momentum distributions.
The~concept of transverse-momentum phase-space is followed,
with a~leading particle effect, in a~similar manner to
ARC~\cite{pan93},
\be
\gamma \, I \propto \prod_{j=1}^N {d {\bf p}_j' \over
\gamma_j'} \, {\rm
e}^{-B \, E_{\perp j}'} \, W_{\parallel j} \times \delta \left
(p_1 + p_2 - \sum_{j=1}^N p_j' \right) \, .
\label{i=}
\ee
The~longitudinal weight is $W_\parallel = {\rm e}^{-|y - y_i|}$
for leading particles and $W_\parallel = 1$
for central ones.

As to low-energy processes, we ensure that, besides elastic, we
include all those needed for the chemical equilibration, i.e.\
$\pi + N \leftrightarrow \Delta$, $\pi + \Delta
\leftrightarrow N + \rho$,
$\pi + \pi \leftrightarrow \rho$,
$\pi + \pi \leftrightarrow \rho + \rho$, $N + N \leftrightarrow
N + \Delta$,
$N + N \leftrightarrow \Delta + \Delta$,
$N + \Delta \leftrightarrow \Delta + \Delta$,
$B + \overline{B} \leftrightarrow \pi + \pi$,
$B + \overline{B} \leftrightarrow \rho + \rho$, and
$B + \overline{B} \leftrightarrow \rho + \pi$.
The~practical implementation of all the processes, though, is
still not completed.

\begin{figure}
\centerline{\includegraphics[angle=-90,
width=.97\linewidth]{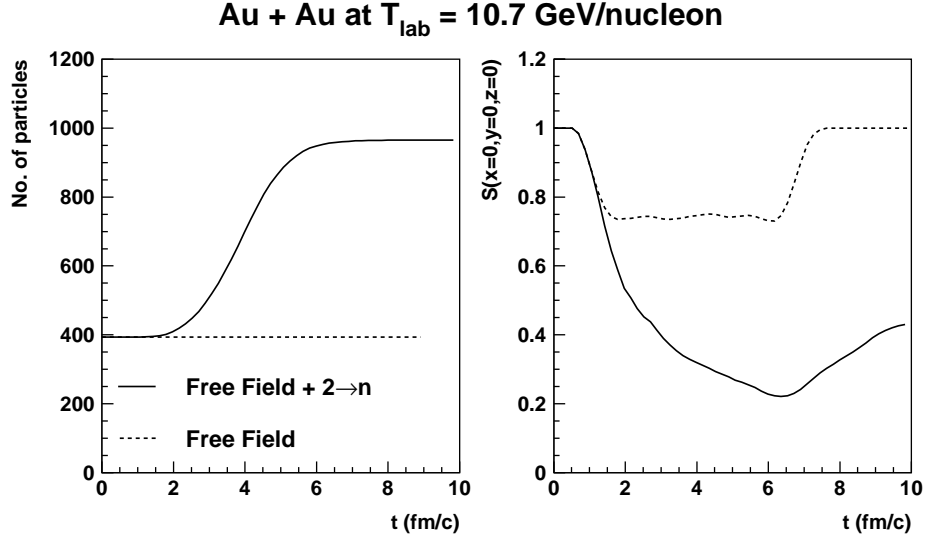}}

\caption{Evolution of the number of hadrons (left) and of
the mass factor~$S$ at the system center (right)
in $b=0$ Au + Au reaction at 10.7~GeV/nucleon.
Dashed lines show the evolution from the transport equation
with mean field only and solid lines show evolution from the
transport equation from the equation with mean field and
collisions.}

\label{Scen}
\end{figure}

With only the high-energy processes in the model, we test
whether the phase
transition may be crossed in the heavy-ion reactions at AGS.
Specifically, we examine the head-on reaction of Au + Au at
10.7~GeV/nucleon.
Figure~\ref{Scen}
shows, as a~function of time, the net
particle number and the mass reduction factor~$S$ at the
center of that system evolved with the mean field only in the
transport equation and with the mean field and the elementary
collisions.  In~the case of mean field only, the~masses drop to
values comparable to those in normal nuclei and then recover.
In~the case with collisions, the~particle number increases by
a~factor of 2.5 compared to the initial state and masses at the
center drop to values such as behind the transition in our
model.  After $t \sim 6$~fm/c in the latter evolution,
the~system appears to fragment into domains of low values
of~$S$ surrounded by regions of values close to~1.

As the QGP phase transition appears to be crossed at AGS
energies, we now turn to the experimental observables that
could signal the crossing.

\section{Elliptic Flow}

Regions of transitional behavior and of phase transitions are
commonly characterized by the changes in the speed of sound.
At~the first-order phase transition, such as in Figs.~\ref{ep}
and~\ref{T0}, the~speed of sound, $c_s = \sqrt{dp/de}$,
vanishes.  Above the phase transition in Fig.~\ref{ep},
the~speed of sound remains low, due to the~slower rise of
pressure~$p$ with temperature than the rise of energy~$e$.
Also
for the situation in Fig.~\ref{T0}, the~speed of sound is low
above the phase transition compared to the region below the
transition.

\begin{figure}
\centerline{\includegraphics[angle=0,
width=.45\linewidth]{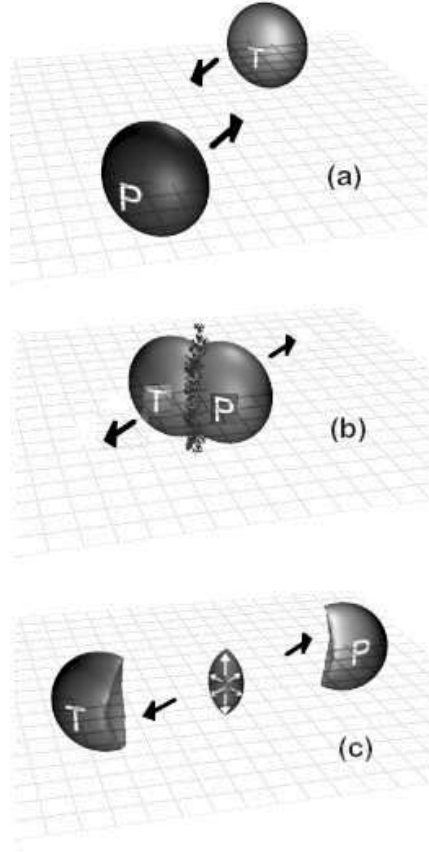}}
\caption{
                Schematic illustration of the collision of
two Au nuclei at relativistic energies.   Time shots are shown for an~instant
before the collision~(a), early in the collision~(b), and late in the
collision~(c).}
\label{fig1}
\end{figure}

A~sensitive measure of the speed of sound or pressure compared
to the energy density early on in the reactions is the elliptic
flow.  The~elliptic flow is the anisotropy of transverse
emission at midrapidity.  At AGS energies, the~elliptic flow
results from a~strong competition~\cite{sor97} between
squeeze-out
and in-plane flow, as illustrated in~Fig.~\ref{fig1}.
In~the early stages of the collision as shown
in Fig.~\ref{fig1}(b), the spectator nucleons block the path of participant
hadrons emitted toward the reaction plane;  therefore the nuclear matter is
initially squeezed out preferentially orthogonal to the reaction plane. This
squeeze-out of nuclear matter leads to negative elliptic flow.  In the later
stages of the reaction, as shown in Fig.~\ref{fig1}(c),
the~geometry of the participant region (i.e.~a~larger
surface area exposed in the direction of the reaction plane)  favors in-plane
preferential emission and hence positive elliptic flow.

The squeeze-out contribution to the elliptic flow and the
resulting net
sign of the flow depend on two factors: (i)~the pressure built
up in the compression stage compared to the energy density, and
(ii)~the passage time
for removal of the shadowning due to the projectile and target spectators.
In~the hydrodynamic limit, the~characteristic time for the
development of expansion
perpendicular to the reaction plane is $\sim R/c_s$, where the speed of sound
is $c_s =\sqrt{\partial p / \partial e}$, $R$~is the nuclear
radius, $p$ is the
pressure and $e$ is the energy density. The~passage time is
$\sim 2R /(\gamma_0
\, v_0)$, where $v_0$ is the c.m.\ spectator velocity.
The~squeeze-out
contribution should then reflect the ratio~\cite{dan95}
\begin{equation}
{c_s \over \gamma_0
\, v_0 } \, .
\label{ratio}
\end{equation}

According to~(\ref{ratio}) the squeeze-out contribution should
drop with the increase in energy, because of the rise in~$v_0$
and then in~$\gamma_0$.  A~stiffer EOS should yield a~higher
squeeze-out contribution.  A~rapid change in the stiffness
with baryon density and/or excitation energy should be
reflected in a~rapid change in the elliptic flow excitation
function.  A~convenient measure of the elliptic flow is the
Fourier coefficient
$\langle
\cos{2 \phi} \rangle \equiv v_2$, where $\phi$ is the azimuthal
angle of a~baryon at midrapidity, relative to the reaction
plane.  When squeeze-out dominates, the~Fourier coefficient is
negative.  When late-stage in-plane emission dominates,
the~coefficient is positive.

\begin{figure}

\centerline{\includegraphics[angle=0,
width=.74\linewidth]{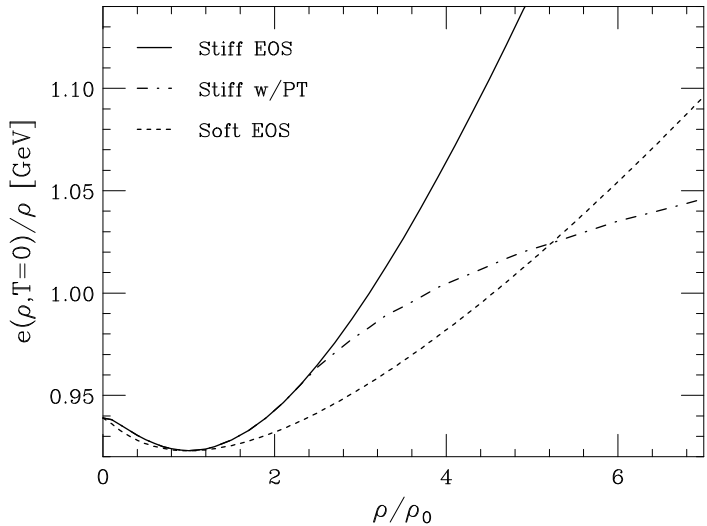}}
\caption{
                Energy per baryon vs.\  baryon density, at $T = 0$. Curves
are shown for a~stiff EOS (solid curve), a~soft EOS
(dashed),
and an~EOS with a~second-order phase transition (dashed-dot).
}

\label{fig2}
\end{figure}

\begin{figure}

\centerline{\includegraphics[angle=0,
width=.87\linewidth]{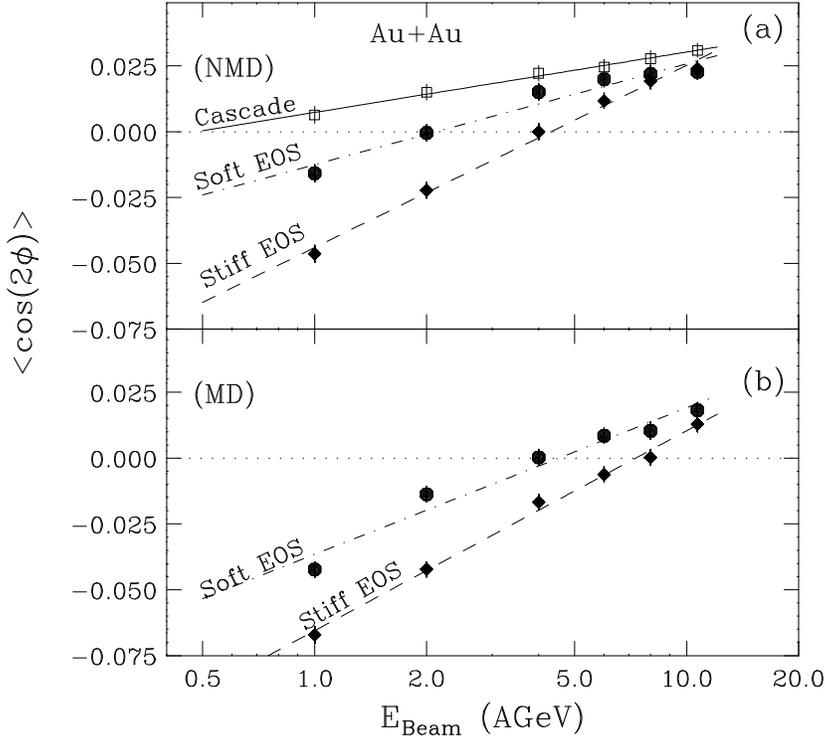}}

\caption{
                        Calculated elliptic flow excitation functions for Au +
Au reactions. Panels~(a) and~(b) show, respectively, the~functions obtained
without~(NMD) and with~(MD) the momentum dependent forces.
The~filled circles, filled diamonds, and open squares indicate,
respectively, results obtained using a~soft EOS, a~stiff EOS,
and by neglecting the mean field.  The~straight lines show logarithmic
fits.
}
\label{fig3}
\end{figure}

To verify whether the expectations regarding the elliptic flow
are
realistic, we have carried out calculations~\cite{dan98} within
a~limited version of the~transport model based on
Eqs.~(\ref{Bol}) and~(\ref{I=}), with nucleon, pion, delta,
and~$N^*$ degrees of freedom.  The~mean fields within the
model acted on baryons only,
giving stiff and soft EOS, at~$T=0$, such as indicated in
Fig.~\ref{fig2}.
We~have also carried out the calculations without mean fields.
The~obtained elliptic-flow excitation functions for Au + Au
reactions, at $b = 4 - 6$~fm, are shown in Fig.~\ref{fig3}(a).
The~elliptic flow changes sign from negative
values
at low beam energies ($ \sim $0.2~AGeV) to positive values at
high beam energies ($ \sim $20~AGeV), as expected from the
weakening of the squeeze-out contribution to the flow.
The~energy
at which the flow sign changes strongly
depends on~EOS.  Over the range
$0.5
\lesssim E_{Beam} \lesssim 10$~AGeV,
the~dependence on the beam energy is essentially logarithmic.
The~slope in Fig.~\ref{fig3}(a) strongly depends on EOS
in~accordance with the considerations involving~(\ref{ratio}).
This raises hope that any change in the stiffness of the~EOS
with energy would be manifested as a~change in the slope of
the excitation function.

Experience from the low-energy domain ($E_{Beam} \lesssim
1$~AGeV) has
shown that the~momentum dependence of the mean field is
an~important factor for
the determination of the stiffness of the~EOS from flow
measurements.
Nucleon-nucleus scattering experiments and nuclear-matter
calculations clearly
indicate the presence of this momentum dependence; in
reactions, it plays a~role in
generating flow before the hadronic matter equilibrates.
A~priori, there could
be two separate effects due to the momentum dependence: (i)~it
may pass for
an~enhanced stiffness of the~EOS, and/or (ii)~it may lead to
a~loss of
sensitivity to the stiffness of the~EOS .  The~second effect
could eliminate
the possibility of observing a~change in the stiffness with
increasing beam energy.

        In Fig.~\ref{fig3}(b), we show elliptic flow excitation
functions
obtained from calculations that include momentum-dependent
fields acting on the baryons~\cite{dan98}.  The~general
trends of these excitation functions are similar to those shown
in
Fig.~\ref{fig3}(a). However, one can clearly see that the net
effect of the
momentum dependence is to enhance the squeeze-out.  Of greater
significance is
the fact that the~sensitivity of the elliptic flow to the
stiffness of the~EOS
remains practically unchanged when this momentum dependence is
included. A~cursory examination of
Fig.~\ref{fig3} also shows that measurements of elliptic flow
should clearly
discriminate between models with a~realistic EOS and the
cascade model in which the elliptic flow becomes positive
right above $\sim 0.6$~AGeV.

\begin{figure}
\centerline{\includegraphics[angle=0,
width=.8\linewidth]{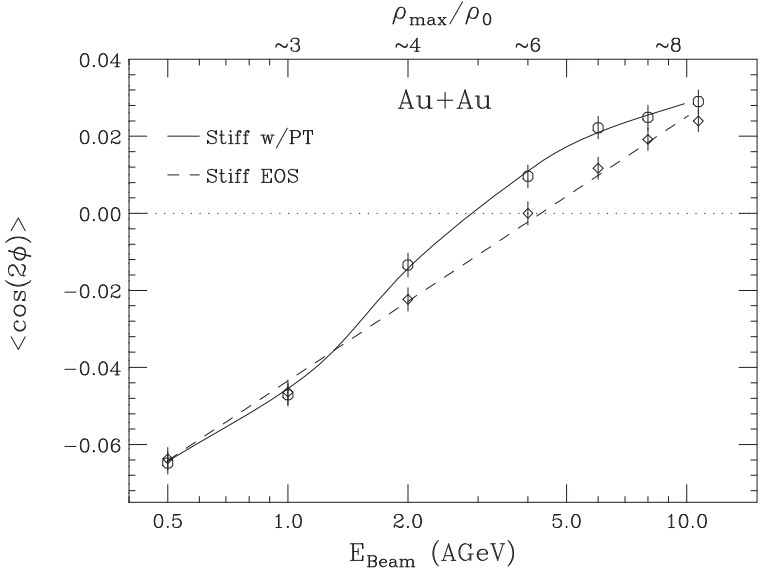}}
\caption{
                  Calculated elliptic flow excitation functions for Au +
Au. The~open diamonds represent results obtained with a~stiff EOS. The~open
circles represent results obtained with a~stiff EOS  and with a~second-order
phase transition.  The~solid and dashed lines are drawn to
guide the eye.  Numbers at the top of the figure indicate rough magnitude
of local baryon densities reached in reactions at different beam energies.
}
\label{fig4}
\end{figure}

        In~order to search for an elliptic-flow signature that
can signal the
onset of a~phase transition to the~QGP, we have carried out
calculations
assuming a~stiff EOS with a~weak second-order phase transition
at $\rho_v = 2.3 \, \rho_0$, indicated by the dash-dot line in
Fig.~\ref{fig2}.
The~elliptic-flow excitation functions calculated using
a stiff EOS with the phase transition (open circles) and
a~stiff EOS without the
phase-transition (diamonds) are compared in Fig.~\ref{fig4}.
Both functions
have been obtained with no consideration of momentum dependence
in the mean
field.  For low beam energies ($\lesssim \, 1$~AGeV),
the~elliptic flow
excitation functions are essentially identical because the two
EOS are either
identical or not very different at the densities and
temperatures that are
reached.  For $2 \lesssim E_{Beam} \lesssim 9$~AGeV
the~excitation function
shows  larger in-plane elliptic flow from the calculation which
includes the
phase transition, indicating that a~softening of the EOS has
occured for this beam energy range.
This deviation is in direct contrast to the esentially
logarithmic beam energy
dependence obtained (for the same energy range) from the
calculations which
assume a~stiff EOS without the phase transition.
Collection of the data on elliptic flow from EOS, E895, and
E877 Collaborations~\cite{pin98} points to a~softening at
a~somewhat higher beam energy, $\sim 3$~AGeV, than in the
reported
transport calculations, corresponding to the baryon density
$\rho_v \sim 4 \, \rho_0$.

\section{Conclusions}

To conclude,
we have specified a~tractable transport model with
thermodynamic properties close to those known or expected for
the strongly
interacting matter.  In~the phase transition region in the
model, along the $\mu =0$ axis, the~masses of elementary
excitations drop while the number of active degrees of freedom
increases.  We have shown that the elliptic flow measurements
can discriminate between models with nontrivial EOS and cascade
models.  The~data on the elliptic flow excitation function
point to the softening of the EOS in the baryon density region
of $\sim 4\, \rho_0$ ($E_{Beam} \sim 3$~AGeV).  These results
show a~definite need to continue the 1--10~AGeV
physics~(SIS-AGS).
This is the region where crossing of the phase transition may
be observed in the excitation functions.  These should be
established with the best possible accuracy.

\section*{Acknowledgements}
        This work was supported in part
by the
National Science Foundation under Grant No.\ PHY-96-05207
and
by the U.S.\ Department of
Energy under Contract No.\ DE-FG02-87ER40331.A008.

\end{document}